\documentclass[showpacs,preprintnumbers,prd,amsmath,amssymb]{revtex4}
\usepackage{amsmath}
\usepackage{graphicx}
\usepackage{subfig}
\usepackage{hyperref}
\usepackage{amssymb}
\usepackage[english]{babel}
\usepackage{epsfig}
\usepackage{wasysym}

\begin{document}

\title{Regular black holes in $f(R)$ gravity coupled to nonlinear
electrodynamics\\}

\author{ Manuel E. Rodrigues$^{(a,c)}$\footnote{E-mail address: 
esialg@gmail.com},  Ednaldo L. B. Junior$^{(a,b)}$\footnote{E-mail 
address: ednaldobarrosjr@gmail.com}, Glauber T. 
Marques$^{(d)}$\footnote{E-mail 
address: gtadaiesky@hotmail.com} and Vilson T. Zanchin$^{(e)}$\footnote{E-mail 
address: zanchin@ufabc.edu.br}}
\affiliation{$^{(a)}$Faculdade de Ci\^{e}ncias Exatas e Tecnologia, 
Universidade Federal do Par\'{a}\\
Campus Universit\'{a}rio de Abaetetuba, 68440-000, Abaetetuba, Par\'{a}, 
Brazil\\
$^{(b)}$Faculdade de F\'{\i}sica, Programa de P\'os-Gradua\c{c}\~ao em 
F\'isica, Universidade Federal do 
 Par\'{a}, 66075-110, Bel\'{e}m, Par\'{a}, Brazil\\
$^{(c)}$Faculdade de Engenharia da Computa\c{c}\~{a}o, 
Universidade Federal do Par\'{a}, Campus Universit\'{a}rio de Tucuru\'{\i}, 
68464-000, Tucuru\'{\i}, Par\'{a}, Brazil\\
$^{(d)}$ Laborat\'orio de Sistemas Ciberf\'\i sicos, Instituto Ciber 
Espacial,  Universidade Federal Rural da Amaz\^{o}nia\\
Avenida Presidente Tancredo Neves 2501,  66077-901, Bel\'{e}m, Par\'a, 
Brazil\\
$^{(e)}$Centro de Ci\^{e}ncias Naturais e Humanas, 
Universidade Federal do ABC\\
Avenida dos Estados 5001, 09210-580, Santo Andr\'{e}, S\~{a}o Paulo, Brazil}


\begin{abstract}
We obtain a class of regular black hole 
solutions in four-dimensional $f(R)$ gravity, $R$ being the curvature 
scalar, coupled to a nonlinear electromagnetic source. The metric formalism 
is used and static spherically symmetric spacetimes are assumed.
The resulting $f(R)$ and nonlinear electrodynamics functions are
characterized by a one-parameter family of 
solutions which are generalizations of known regular black holes in 
general relativity coupled to nonlinear electrodynamics. The related
regular black holes of general relativity 
are recovered when the free parameter vanishes, in which case one has 
$f(R)\propto R$.
We analyze the regularity of the solutions and also show that there are
particular solutions that violate only the strong 
energy condition 
\end{abstract}

\date{\today}

\maketitle



\section{Introduction}
\label{sec1}

The theory of general relativity (GR) has passed many experimental and
observational tests and is widely accepted as the best theory of
gravitation. Even with such a success, there are some aspects  in GR to be
better understood. In this sense, the prediction of spacetime singularities 
and the difficulty to be quantized
through standard methods are two important issues to be mentioned.
The singularity problem is connected to the existence of black holes, which 
are probably
the most intriguing objects predicted by GR. In the simplest case,
the Schwarzschild black hole, 
the spacetime presents an event horizon which hides a curvature singularity 
inside of it. More generally, if the matter-energy content of the spacetime 
satisfies some reasonable physical conditions, the so-called energy 
conditions, the singularity theorems of GR \cite{sing-theo} assure that the 
singularity is inevitable. 
Even though the singularity can be hidden by the presence of an event 
horizon, which protects the exterior world, its own presence signals the 
breakdown of the 
physical laws. In other words, a good physical theory should be free 
of singularities. It is believed that a complete theory of quantum gravity 
would have such a good property, but such a theory does not exist yet. 
Hence, while a theory of quantum gravity is not formulated, good strategies 
to get rid of singularities is to search for alternative matter-field models 
within GR, and to search for alternative theories of gravity that are free 
of singularities.

Within GR, singularities can 
be avoided if the energy conditions assumed by the singularity theorems 
are somehow violated. In the case of black holes, the singularity can be 
replaced by a regular region filled by some kind of matter or field that 
violates at least the strong energy condition. In fact, the
matter content of the first regular black hole solution found in GR 
is such that, near the center, it satisfies a de Sitter type equation of 
state $p= -\rho$, with $\rho$ and $p$ being the energy density and the 
pressure of an effective perfect fluid model representing the energy-momentum 
tensor of the Bardeen black hole \cite{Bardeen}. It clearly  violates the 
strong energy condition. As a matter of fact, a
perfect fluid satisfying a de Sitter equation of state has long been
suggested to solve the cosmological singularity problem
\cite{sakharov-gliner}. A wide class of regular black holes
solutions found in the literature follows this idea, the singularity being
replaced by a regular distribution of matter satisfying a de Sitter type
equation of state at least at the central core. The class of solutions that 
followed Bardeen's idea evolved towards models whose source is the 
energy-momentum coming
from some kind of nonlinear electromagnetic theory (see, e.g.,
\cite{ABG-ned}, see \cite{reviewRBH} for a review). A second very 
interesting class of
regular black holes followed precisely the idea of modeling the matter
content by a perfect fluid obeying the relation $p=-\rho$ (see, e.g.,
\cite{dym-deSitter}).

 The singularity problem in GR is linked to the important observational 
result that the Universe is in a phase of accelerated expansion, as first 
inferred from 
type Ia supernovae data \cite{SNIa} and confirmed by several other tests 
(see, e.g., \cite{accel} for reviews). Considering 
a Friedman-Lama\^itre-Robertson-Walker cosmological spacetime fulfilled by 
a perfect fluid, in order to have an accelerated expansion, Einstein equations 
require the perfect fluid to satisfy the 
condition $\rho + 3p < 0$, where $\rho$ and $p$ are respectively the energy 
density and pressure of the fluid. More than that, in the 
cosmic concordance model, namely, the $\Lambda$CDM model, the 
perfect fluid model for the dark energy satisfies $p=-\rho$. Assuming an 
equation of state in the form $p = \omega \rho$,  several 
cosmological tests indicate that the present 
accelerated phase of the Universe requires $\omega \lesssim -1$ \cite{lima}.
These facts motivated more studies on regular black holes modeled by de 
Sitter
like and phantom matter fields \cite{deSitter2-phantom} (see also 
\cite{dym-deSitter}).

Beyond GR there are many proposals of alternative theories of gravity. 
Besides trying to solve the singularity problem and to avoid 
introducing matter with nonstandard physical properties, alternative
theories have been proposed to account for quantum corrections in the
action. A well established proposal is the $f(R)$ gravity 
\cite{fR,deFelice} which replaces the Einstein-Hilbert Lagrangian density 
by a general function of the Ricci scalar $R$. The already mentioned issue 
of an accelerated cosmological expansion phase motivated also several other 
proposals 
which are still being developed. Worth of mentioning are the $f(R,\Theta)$ 
\cite{fRT}, the $f(G)$ \cite{fG}, the $f(R,G)$ \cite{fRG}, and the 
$f(R,\Theta,R_{\mu\nu}\Theta^{\mu\nu})$ \cite{odintsov} gravity theories. 
In this notation, $G$ stands for the Gauss-Bonnet term,  
$\Theta^{\mu\nu}$ and $\Theta$ represent the energy-momentum 
tensor and its trace, respectively, and $R^{\mu\nu}$ is the Ricci 
tensor. There are many applications of such generalized $f(R)$ 
gravity theories in cosmological models (see the reviews of Refs. 
\cite{fR,deFelice}, see also \cite{fR-cosmol}).  
 Similarly, in the context of the Teleparallel Theory of gravity 
(TTG) \cite{TT}
there are also analogous models that modify the standard TTG into more 
general theories. The $f(T)$ theory is the analogous of the $f(R)$ gravity
whose modification was introduced to try to account for the same 
problems of GR (see \cite{fT} for a list of references). Further 
generalizations are the $f(T,\Theta)$ \cite{fTTc}, 
the $f(T,T_G)$ \cite{fTTG}, and the $f(\mathcal{T})$ gravity theories 
\cite{ednaldo}. For a review on the applications of modified theories of 
gravity see, e.g., Ref. \cite{bamba} (see also \cite{fT-cosmol}).

On the other side, considering matter and fields models, the Maxwell 
electromagnetism is also a very well established theory that has passed all 
the experimental tests. But, with 
the aim of avoiding divergent terms in quantum electrodynamics, 
Born and Infeld \cite{BI} proposed a modified version of the Maxwell theory 
which, together with other generalizations, is dubbed the nonlinear 
electrodynamics (NED) \cite{peres}. When coupled to GR, 
the NED theory produces regular black hole solutions where the singularity 
is avoided by a regular distribution of the (new) electromagnetic field that 
fulfills the central core of the black hole. In fact, construction of regular 
solutions of GR coupled to NED has been largely studied in the 
literature \cite{NED}. This fact motivates 
the construction of regular black hole solutions in the new (generalized) 
gravity theories coupled to the generalized models for matter fields, since it
is a natural way to investigate the properties of the new theories in the strong 
field regime.

Within $f(R)$ gravity theories many black hole solutions are found 
in the literature, either using the metric formalism \cite{metricfR} 
or the Palatini formalism \cite{PalatinifR}. Yet, in order to 
find regular black hole solutions, special kinds of matter are needed. For 
instance, nonlinear electromagnetic and Yang-Mills fields are considered 
as sources to build regular $f(R)$ gravity black holes. In other cases, 
including electric charge, the matter distribution must satisfy particular 
conditions, generating a wormhole type structure avoiding thus the 
central singularity.

Motivated by the scenario summarized above, in the present work we follow 
the idea of coupling a nonlinear electrodynamics (NED) to the chosen (new)
gravity theory, and establishing a method to find exact solutions in 
such a way to be able to built regular black hole solutions within such a 
theory. 

The paper is structured as follows. In Sec. \ref{sec2} the basic 
equations for the $f(R)$ theory minimally coupled to a general 
electromagnetic action are presented, a spherically symmetric static
metric is chosen and the explicit field equations for the metric functions 
and electromagnetic field are given in terms of a Schwarzschild radial 
coordinate $r$. Section \ref{sec3} is devoted to present new regular black 
holes solutions, it is divided into four subsections. In the first 
subsection, \ref{sec3.1}, after making a simplifying hypothesis, the only
relevant metric coefficient is written in terms of a generic mass function 
$M(r)$. Function $f_R$, that is the derivative of $f(R)$ with respect to 
$R$, is integrated 
resulting in a linear function of the radial coordinate. All the relevant 
equations are then written in terms of $M(r)$ and the energy conditions are 
written explicitly in terms of the effective energy-momentum tensor of the 
theory. 
In subsection \ref{sec3.2}, we show that the ansatz for $M(r)$ reproduces the 
Reissner-Nordstr\"om black hole solution in the particular case where 
$f_R=1$, verifying the consistency of the procedure used to generate the
exact solutions. Subsection \ref{sec3.3} is reserved to present a 
solution which has two horizons and is regular throughout the spacetime. 
A particular form of $M(r)$ is chosen in order to produce regular black holes, 
as known from previous works. The solution is new in the sense that it is the
first of this class within $f(R)$ gravity coupled to NED theories.
The 
solution is analyzed in some detail and it is shown that the weak energy 
condition is violated in the central region of the spacetime inside the 
event horizon. A careful analysis of the regularity and asymptotic behavior 
of this solution is performed in Sec.~\ref{secA1} of Appendix \ref{secA}.
In subsection \ref{sec3.4}, a second solution representing a regular black 
hole in $f(R)$ coupled to NED theory is presented and analyzed.
In this case, only the strong energy 
condition is violated inside the event horizon. The regularity of the 
solution at the asymptotic limit (spatial infinity) of this solution is 
considered in Sec.~\ref{secA2} of Appendix \ref{secA}.
In Sec. \ref{sec4} we make final remarks and conclude.

\section{The equations of motion in $f(R)$ gravity}
\label{sec2}

The $f(R)$ gravity is a generalization of GR which in the metric formalism 
is obtained from the action (we follow the sign conventions of 
Ref.~\cite{deFelice}).
\begin{eqnarray}
S_{f(R)}=\int d^4x\sqrt{-g}\left[f(R)+2\kappa^2\mathcal{L}_m\right]
\label{action},
\end{eqnarray}
where $g$ stands for the determinant of the metric $g_{\mu\nu}$, $f(R)$ is a 
given function of the Ricci scalar $R$, $\mathcal{L}_m$ represents the 
Lagrangian density of the matter and other fields, and 
$\kappa^2=8\pi G/c^4$, with $G$ and $c$ being the Newton's 
gravitational constant and the speed of light, respectively (from now on we  
choose units such that $G=1$ and $c=1$). Varying action \eqref{action} with 
respect to the metric it results
\begin{eqnarray}
f_RR^{\mu}_{\;\;\nu}-\frac{1}{2}\delta^{\mu}_{\nu}f+\left(\delta^{\mu}_{\nu} 
\square 
-g^{\mu\beta}\nabla_{\beta}\nabla_{\nu}\right)f_R=\kappa^2\Theta^{\mu}_{
\;\;\nu}\label{eqfR}\;,
\end{eqnarray}  
where $f_R\equiv df(R)/dR$, $R^{\mu}_{\;\;\nu}$  is the Ricci 
tensor, $\nabla_{\nu}$ stands for the (metric compatible) covariant 
derivative, $\square\equiv 
g^{\alpha\beta}\nabla_{\alpha}\nabla_{\beta}$ is the d'Alembertian, and 
$\Theta_{\mu\nu}$ is the matter energy-momentum tensor.

In the present work we consider a nonlinear electrodynamics (NED) model 
coupled to the $f(R)$ gravity, so that the Lagrangian density for the matter 
fields in the action (\ref{action}) may be particularized as 
$\mathcal{L}_m\equiv 
\mathcal{L}_{NED}(F)$, where $F=(1/4)F^{\mu\nu}F_{\mu\nu}$, and with  
$F_{\mu\nu}$ being the Faraday-Maxwell tensor (here we neglect the 
electromagnetic sources). In such a case, the 
energy-momentum tensor is written as
\begin{eqnarray}
&&\Theta^{\mu}_{\;\;\nu}=\delta^{\mu}_{\nu}\mathcal{L}_{NED}-\frac{
\partial\mathcal{L}_{NED}(F)}
{\partial F}F^{\mu\alpha}F_{\nu\alpha}\label{energy}\;.
\end{eqnarray}
The canonical energy-momentum tensor of the Maxwell theory follows in the 
particular case where $\mathcal{L}_{NED}\equiv F$.

The Faraday-Maxwell tensor may be given in terms of a gauge potential in the 
usual form 
$F_{\mu\nu}=\partial_{\mu}A_{\nu}-\partial_{\nu}A_{\mu}$. Then, varying 
action \eqref{action} with respect to the potential $A_{\mu}$ one gets a 
generalized version of the Maxwell equations
\begin{eqnarray}
\nabla_{\mu}\left[F^{\mu\nu}\mathcal{L}_F\right]\equiv  
\partial_{\mu}\left[\sqrt{-g}F^{\mu\nu}\mathcal{L}_F\right]=0
\label{Maxwell}\;,
\end{eqnarray}
where $\mathcal{L}_F=\partial\mathcal{L}_{NED}/\partial F$.

Furthermore, taking the trace of Eq.~\eqref{eqfR} one 
gets   
\begin{eqnarray}
f_RR-2f+3\square f_R=\kappa^2 \Theta\label{constraint}\,,
\end{eqnarray}
where $\Theta\equiv g^{\mu\nu}\Theta_{\mu\nu}=4\mathcal{L}_{NED} 
-F\mathcal{L}_F$. 

We then consider a spherically symmetric and static spacetime, whose 
line element, in Schwarzschild-like coordinates, reads
\begin{eqnarray}
ds^2=e^{a(r)}dt^2-e^{b(r)}dr^2-r^2\left[d\theta^2+\sin^2\theta 
d\phi^2\right] 
\label{ele}\;,
\end{eqnarray} 
where $a(r)$ and $b(r)$ are arbitrary functions of the radial coordinate 
$r$ alone. With this, the electromagnetic source is the electric charge 
only, and the magnetic parts of the Faraday-Maxwell tensor $F_{\mu\nu}$  
vanish identically. Moreover, due to the symmetry the only nonzero component 
is electric intensity $F^{10}(r)$ \cite{wainwright}. Hence, we can integrate
the resulting equation from  \eqref{Maxwell} for $\nu =0$ to get,
\begin{eqnarray}
F^{10}(r)=\frac{q}{r^2}e^{-\left(a(r)+b(r)\right)/2}\mathcal{L}_F^{-1}
(r) \label {F10-1} \; ,
\end{eqnarray} 
where $q$ is an integration constant representing the electric 
charge of the source.

The equations of motion for the $f(R)$ gravity coupled to a NED are then 
found by using the line element \eqref{ele}, the Faraday-Maxwell tensor 
\eqref{F10-1}, the energy-momentum \eqref{energy}, and the equations of 
motion 
\eqref{eqfR},
\begin{eqnarray}
&&\frac{e^{-b}}{4r}\Big\{4r\frac{d^2f_R}{dr^2}+ 
2\left[4-rb'\right]\frac{df_R}{dr}+\big[ra'b'-2ra''-r(a')^2-4a'\big]
f_R+2re^bf\Big\}\nonumber\\
&&\qquad =-\kappa^2\left[\mathcal{L}_{NED}+ 
\frac{q^2}{r^4}\mathcal{L}_F^{-1}\right] ,\label{eq1}\\
&& \frac{e^{-b}}{4r}\Big\{2\left[4+ra'\right]\frac{df_R}{dr}+ 
\left[(4+ra')b'-2ra''-r(a')^2\right]f_R +2re^bf\Big\} \nonumber\\
&&\qquad =-\kappa^2\left[\mathcal {L}_{NED}+
\frac{q^2}{r^4}\mathcal{L}_F^{-1}\right],\label{eq2}\\
&&\frac{e^{-b}}{2r^2}\Big\{2r^2\frac{d^2f_R}{dr^2}+[r^2(a'-b')+2r]
\frac{df_R}{dr}+[r(b'-a')+2(e^b-1)]f_R+r^2e^bf\Big\} 
\nonumber\\
&&\qquad =-\kappa^2\mathcal{L}_{NED} ,\label{eq3}
\end{eqnarray}
where the prime ($'$) stands for the total derivative with respect to the 
radial coordinate $r$. 

To be complete, we stress that Eq. \eqref{constraint} is satisfied by 
any solution of Eqs.~\eqref{eq1}-\eqref{eq3}. This can be shown explicitly 
by combining these three equations and noticing that the Ricci scalar for 
the present case is given by 
\begin{eqnarray}
R=e^{-b}\left[a''+\left(a'-b'\right)\left( \frac{a'}{2} +\frac{2}{r}\right)
+\frac{ 2}{r^2}  \right]-\frac{2}{r^2}\label{R}\;.
\end{eqnarray}

In the next section we shall solve of Eqs. 
(\ref{eq1})-(\ref{eq3}) and present the first solutions of regular black 
holes in the $f(R)$ gravity coupled to a nonlinear electrodynamics theory.

\section{New  regular black hole solutions}\label{sec3}

\subsection{Simplifying hypothesis and the ansatz for the mass function}
\label{sec3.1}
Once the spacetime geometry is assumed to be static and to carry the 
spherical symmetry, all the metric function and fields depend on the radial
coordinate only and then the system of differential equations may be further
simplified by making suitable hypotheses. Now, by subtracting 
Eq.~\eqref{eq2} from Eq.~\eqref{eq1} it results 
\begin{eqnarray}
\frac{e^{-b}}{2r}\left[2r\frac{d^2f_R}{dr^2}-
\left(r\frac{df_R}{dr}+2f_R\right)\left(a'+b'\right)\right]=0.\label{eq4-0}
\end{eqnarray}
This is a second-order ordinary differential equation for $f_R$, whose 
general solution can be found once functions $a(r)$ and $b(r)$ are known.
However,
these two function are found in general just by solving the complete
nonlinear system of coupled equations resulting from Eqs.~\eqref{eq1},
\eqref{eq2}, and \eqref{eq3} after specifying the NED Lagrangian. 
Since we are interested in black hole solutions, the usual strategy at this
point is to make an ansatz that leads to that kind of solutions. 
An interesting choice is to fix the following relation
 \begin{eqnarray}
b(r)=-a(r) \label{bconstraint}.
\end{eqnarray}
Such a relation is a consequence of the field equations in GR, but that is 
not the case in $f(R)$ gravity theories, where such a choice represents a further 
constraint on the possible solutions. By fixing the metric coefficients as 
given in~\eqref{bconstraint},  equation \eqref{eq4-0} becomes  
\begin{eqnarray}
e^{-b(r)}\frac{d^2f_R}{dr^2}=0\label{eq4}\;.
\end{eqnarray}
The general solution of such an equation, for $e^{-b(r)}\neq 0 $, is given 
by the expression
\begin{eqnarray}
f_R(r)=c_1r+c_0\, ,\label{fR1}
\end{eqnarray}
where $c_1$ and $c_0 $ are integration constants.
Within the class of spherically symmetric systems, this approach has
recurrently been used to find new solution of $f(R)$ gravity theories (see, e.g.,
Eq.~(12) of \cite{cognola} and Eq.~(15) of \cite{sebastiani}).  
Here it is then seen that the solution \eqref{fR1} leads to a simple 
generalization of GR theory. In the 
case with $c_1=0$ and $c_0=1$, we recover GR, since the integration of
Eq.~\eqref{fR1} would furnish  $f(R)=R$. On the other hand, in the general case 
when $c_1,c_0\, \neq0$ and once the function $a(r)$ and $b(r)$
are given, Eq.~\eqref{R} can be inverted to write 
$r$ as a function of $R$,  $r\equiv r(R)$, and finally the integration 
of \eqref{fR1} yields
\begin{eqnarray}
f(R)=c_0R+c_1\int r(R)dR\label{f1},
\end{eqnarray} 
where we have set an integration constant to zero.
From Eq.~\eqref{f1} it is clearly seen that the second term is the responsible for 
the contributions to the $f(R)$ gravity beyond the standard GR, including 
possibly nonlinear terms on the Ricci scalar $R$. 

The linear dependence of $f_R$ on $r$ in Eq.~\eqref{fR1} implies it is an
unbounded function at the
asymptotic limit $r\rightarrow +\infty$. However, as we show explicitly in
Appendix \ref{secA}, for the particular solutions presented in this work,
such an unbounded value implies no divergence in
any of the physical and geometric quantities of the corresponding
spacetimes.
In fact, all the equations of motions, curvature invariants, and effective
energy-momentum tensor components are
well behaved at the asymptotic limit $r\rightarrow +\infty$.

For the spherical geometry, specially when dealing with asymptotically flat 
spacetimes, it is convenient to introduce a mass function  $M(r)$ through 
the ansatz
\begin{eqnarray}
e^{-b(r)}=1-\frac{2M(r)}{r}\label{b}\;,
\end{eqnarray}
where  $M(r)$ satisfies the 
condition $\lim_{r\rightarrow 0}[M(r)/r]\rightarrow 0$. This is a 
necessary condition in order to find solutions which are regular at the 
center $r=0$. We also assume that $M(r)$ tend to the ADM mass at infinity, 
namely, $\lim_{r\rightarrow +\infty}M(r)=m$, with $m$ being the 
ADM mass. 

In terms of the mass function $M(r)$, the Ricci scalar \eqref{R} assumes 
the form
\begin{eqnarray}
R(r)=-\frac{2}{r^2}\left[2M'(r)+rM''(r)\right]\label{R1}\;. 
\end{eqnarray}

At this point we may solve equations (\ref{eq1})-(\ref{eq3}) for  
$\mathcal{L}_{NED}$ and $\mathcal{L}_{F} $ in terms of $M(r)$ to get
\begin{eqnarray}
&&\mathcal{L}_{NED}=-\frac{1}{2\kappa^2 r^2}\left[r^2f(r) 
+4c_0M'(r)+2c_1r\right]\label{L}\;,\\
&&\mathcal{L}_F=-\kappa^2 \frac{q^2}{r^2}\big[(c_1r+c_0)rM''(r) 
-(2c_0+c_1r)M'(r)-3c_1M(r)+c_1r\big]^{-1}\label{LF}\,.
\end{eqnarray}
Furthermore, since $f_R$ as a function of $r$ is known from Eq.~\eqref{fR1} we 
may obtain $f(r)$ as
\begin{eqnarray}
f(r)=\int f_R(r)\frac{dR(r)}{dr}dr\label{f2}
\end{eqnarray}
with $f_R(r)$ and $R(r)$ being given respectively by Eqs.~(\ref{fR1}) 
and (\ref{R1}).
 
It is seen that all the equations of motion (\ref{eq1})-(\ref{eq3}), 
including the constraint (\ref{constraint}), are identically satisfied by 
taking into account the set of equations (\ref{bconstraint}), (\ref{fR1}),
(\ref{b}), (\ref{R1}), (\ref{L}), (\ref{LF}) and (\ref{f2}). A last point 
to be verified is the relation between the Lagrangian density 
$\mathcal{L}_{NED}$ 
and its derivative $\mathcal{L}_F$. From the definition of 
$\mathcal{L}_F$, one has
\begin{eqnarray}
\mathcal{L}_F=\frac{\partial \mathcal{L}_{NED}}{\partial F}=\frac{\partial 
\mathcal{L}_{NED}}{\partial r}\frac{\partial r}{\partial F}=\frac{\partial 
\mathcal{L}_{NED}}{\partial r}\left(\frac{\partial F}{\partial 
r}\right)^{-1}\;.\label{Lconstraint}
\end{eqnarray}
In order to check if relation \eqref{Lconstraint} is satisfied we first 
determine the solely nonzero component of the Faraday-Maxwell
tensor $F^{10}$, which is interpreted as the electric intensity filed. This
can be done by replacing Eqs.~(\ref{bconstraint}), 
(\ref{b}), and (\ref{LF}) into Eq.~(\ref{F10-1}) so that
\begin{eqnarray}
&&F^{10}(r)=\frac{1}{q\kappa^2}\Big\{3c_1M(r)+(2c_0+c_1r)M'(r)-r\left[
c_1+(c_0+c_1r)M''(r)\right]\Big\}\label{F10-2}\;.
\end{eqnarray}
Now, by using the set of Eqs.~(\ref{bconstraint}), (\ref{fR1}), (\ref{b}), (\ref{R1}), 
 (\ref{L}), (\ref{LF}), (\ref{f2}), and the identity 
$F=(1/4)F^{\mu\nu}F_{\mu\nu}=(-1/2)e^{a+b}\left[F^{10}(r)\right]^2$,  it is 
straightforward to show that the constraint (\ref{Lconstraint}) is 
identically satisfied.

Therefore, a new class of solutions of the $f(R)$ gravity is obtained for 
each given function $M(r)$. In fact, the new class of solutions 
depends solely on the function $M(r)$, which must be properly chosen to 
satisfy some reasonable physical conditions. Below we will present some 
particular choices for $M(r)$ that produce regular black hole solutions of 
the $f(R)$ theory coupled to a NED. For each chosen function it results a
different nonlinear electrodynamics theory and a different $f(R)$ theory.

Among the physical conditions a given solution should satisfy there 
are the so-called energy conditions (EC). In order to formulate such
conditions in an appropriate form,  let us rewrite Eqs.~(\ref{eqfR}) in 
terms an effective energy-momentum $\mathcal{T}_{\mu\nu}^{(eff)}$ as follows,
\begin{eqnarray}
&&\hspace{-.3cm}R_{\mu\nu}-\frac{1}{2}g_{\mu\nu}R=
f_R^{-1}\big[\kappa^2\Theta_ { \mu\nu }
+\frac{1}{2}g_{\mu\nu}\left(f-Rf_R\right) -\left(g_{\mu\nu}\square-\nabla_{
\mu}\nabla_{\nu}\right)f_R\big] \equiv \kappa^ { 2
}  \mathcal{T}_{\mu\nu}^{(eff)}\label{energyeff}.
\end{eqnarray}
As defined here, quantity $\mathcal{T}_{\mu\nu}^{(eff)}$ represents the 
effective 
energy-momentum tensor coming from the $f(R)$ gravity that acts as the 
effective source term in Einstein equations. It contains the canonical 
energy-momentum of matter fields, $\Theta_{\mu\nu}$, weighted by $f_R^{-1}$, 
and the additional contributions due to the presence of the nonlinear  $f(R)$ 
function in the Lagrangian density. Based on such alterations of the source 
terms of Einstein equations, possible changes in the energy conditions when 
compared to GR theory are expected. 

After performing the identifications  
$\mathcal{T}_{0}^{0(eff)}=\rho^{(eff)}$, $\mathcal{T}_{1}^{1(eff)}=-p_{r}^{
(eff)}$, and 
$\mathcal{T}_{2}^{2(eff)}=\mathcal{T}_{3}^{3(eff)}=-p_{t}^{(eff)}$, the 
energy conditions (EC) are given by (see, e.g., \cite{visser} for the EC in 
GR, and Refs.~\cite{santos} for studies related to the EC in
$f(R)$ theories)
\begin{eqnarray}
&&NEC_{1,2}(r)=\rho^{(eff)}+p_{r,t}^{(eff)}\geq 0\;,\label{cond1}\\
&&SEC(r)=\rho_{(eff)}+p_{r}^{(eff)}+2p_{t}^{(eff)}\geq 0\,,\label{sec}\\
&&WEC_{1,2}(r)=\rho^{(eff)}+p_{r,t}^{(eff)}\geq 0\;,\label{cond2}\\
&& DEC_{1}(r)=\rho^{(eff)}\geq 0, \label{cond3a}\\ &&
DEC_{2,3}(r)=\rho^{(eff)}-p_{r,t}^{(eff)}\geq 
0\;,\label{cond3}
\end{eqnarray}
where, in view of the identity $WEC_3(r)\equiv DEC_1(r)$, one of the 
conditions is not written.

From Eqs. (\ref{energyeff}) with metric \eqref{ele} we find the relations
\begin{eqnarray}
&&
\rho^{(eff)}=\frac{e^{-b}}{4\kappa^2r^2f_R}\Bigg\{4\kappa^2r^2\left[F^{10 } 
\right]^2e^{a+2b}\mathcal{L}_F+4\kappa^2r^2e^b\mathcal{L}_{NED} 
+4r^2\frac{d^2f_R}{dr^2} +\left(8r-2r^2b'\right)\frac{df_R}{dr}\nonumber\\
&& \qquad\quad +\big[(r^2a'+4r)b'+4e^b-2r^2a''-r^2\left(a'\right)^2 
-4ra'-4\big] f_R +2r^2e^bf \Bigg\} \label{rhoeff}\;,\\
&& p_{r}^{(eff)}= 
-\frac{e^{-b}}{4\kappa^2r^2f_R}\Bigg\{4\kappa^2r^2\left[F^{10}\right]^2e^
{a+2b}\mathcal{L}_F+4\kappa^2r^2e^b\mathcal{L}_{NED}+(2r^2a'+8r)\frac{df_R}{
dr} +\big[ (r^2a'+4r)b'\nonumber\\
&&\qquad\quad +4e^b-2r^2a''-r^2\left(a'\right)^2-4ra'-4\big]
f_R+2r^2e^bf\Bigg\}\;,\label {preff}\\
&& p_{t}^{(eff)}= 
-\frac{e^{-b}}{4\kappa^2r^2f_R}\Bigg\{4\kappa^2r^2e^b\mathcal{L}_{NED}
+4r^2\frac{d^2f_r}{dr^2}+[2r^2(a'-b')+4r]\frac{df_R}{dr}+\big[
(r^2a'+4r)b'\nonumber\\
&&\qquad\quad  +4e^b-2r^2a''-r^2\left(a'\right)^2-4ra'-4\big]
f_R+2r^2e^bf\Bigg\}\;.\label {pteff}
\end{eqnarray}

\subsection{Reissner-Nordstr\"om solution}
\label{sec3.2}
Let us first check the consistency of the model by assuming that the mass 
function is identical to the case of Reissner-Nordstr\"om metric in GR 
coupled to Maxwell electrodynamics theory, namely,
\begin{eqnarray}
M(r)=m-\frac{q^2}{2r}\label{m1}\;,
\end{eqnarray} 
where $m$ is the ADM mass and $q$ is the electric charge of the source. With 
this assumption nothing changes with respect to relations 
(\ref{bconstraint}) and (\ref{b}) and, moreover, Eq. \eqref{fR1} is still 
the general solution of Eq.~\eqref{eq4}.  However, the choice given by 
\eqref{m1} replaced into Eq.~(\ref{R1}) implies in $R(r)=0$, and therefore 
it cannot be used to obtain the relation $r=r(R)$. Hence, since there is no 
relation between the radial coordinate $r$ and the Ricci scalar $R$, the only
way to satisfy Eq.~\eqref{fR1} is to take $c_1=0$, from what follows
$f_R\equiv c_0$ and, by using Eq.~\eqref{f1}, $f(R)=c_0R$. 
The equations are then 
solved for $\mathcal{L}_{NED}$ and $\mathcal{L}_F$, i.e., 
\begin{eqnarray}
\mathcal{L}_{NED}=-\frac{c_0 q^2}{\kappa^2 r^4},\quad
\mathcal{L}_F=\frac{\kappa^2}{2c_0}\label{LLFRN0}\;.
\end{eqnarray} 
With such results we get $F^{10}(r)=2c_0q/(\kappa^2 r^2)$ and 
$F=-2c_0^2q^2/(\kappa r)^4$. Since now the nonlinear electrodynamics turned 
into the linear Maxwell theory we must have $\mathcal{L}_{NED}=F$, what 
implies $c_0=\kappa^2/2$. At the end we are left with the 
 Reissner-Nordstr\"om solution, i.e., 
\begin{eqnarray}
&&\hspace{-.6cm} a(r)=-b(r)=\log\left[1-\frac{2m}{r}+\frac{q^2}{r^2}\right],
\;\; f(R) =\frac{\kappa^2}{2}R,\\
&&\hspace{-.6cm}f_R=\frac{\kappa^2}{2},\;\; F^{10}=\frac{q}{r^2}, \;\;
\mathcal{L}_{NED}=-\frac{q^2}{2r^4},\;\;\mathcal{L}_F=1.
\end{eqnarray}

Here we have seen that for a given mass function $M(r)$ it follows a 
particular solution. As a consistency check we rebuilt the 
Reissner-Nordstr\"om solution, a solution of the ordinary GR theory coupled 
to Maxwell electrodynamics, which is recovered by using the fact that
for vanishing Ricci scalar the arbitrary parameter $c_1$ has to be set to zero.
This process may be understood by noticing that 
the particular choice of $M(r)$ defines the Ricci scalar and, through 
Eq.~\eqref{L}, it also gives $\mathcal{L}_{NED}(F)$, determining a
particular model of NED theory.
As a matter of fact, the generalization path adopted here restricts the
spacetime geometry to the case where the Ricci scalar $R$ is nonzero, 
so that it is possible to
express the radial coordinate $r$ in terms of the curvature scalar $R$.  
For zero Ricci scalar, the procedure does not apply and alternative routes have to be 
followed. 

In the next two sections we shall see how to obtain solutions for $f(R)$ gravity 
coupled to NED that may be view as generalizations of known solutions within GR theory. 
The solutions reported below are new in the sense that they are the first
regular black hole solutions of the $f(R)$ gravity coupled to a nonlinear
electrodynamics theory.

\subsection{First new regular black hole solution}
\label{sec3.3}
Let us now take the mass function given by
\begin{eqnarray}
M(r)=me^{-q^2/(2mr)}\label{m2},
\end{eqnarray}
where $m$ and $q$ are constant parameters.
This model was considered in Ref.~\cite{coletu}, and was also analyzed 
in Ref.~\cite{balart} (see Table $1$ of such a reference).
From Eqs.~(\ref{m2}), (\ref{bconstraint}) and (\ref{b}), it follows  
\begin{eqnarray}
e^a(r)=e^{-b(r)}=1-\frac{2m}{r}e^{-q^2/(2mr)}\label{a1}\,.
\end{eqnarray}
As well established, the horizons of a metric of the form \eqref{ele} 
are given by the solutions of the equation $g^{11}(r_H)=e^{-b(r_H)}=0$. 
Using the Lambert $W$ function \cite{dence}, which solves the 
equation $W[x]e^{W[x]}=x$, it is found that the horizons radii 
are given by 
$r_H=[-q^2/(2m)]/W[-q^2/(2m)^2]$. In fact, the situation in this case is 
similar to the Reissner-Nordstr\"om spacetime, there are three  
possibilities here: (i) two real and positive roots (nondegenerated case), 
meaning an event and a Cauchy horizon; (ii) one (degenerated) real and 
positive root, the event and the 
Cauchy horizons coincide; and 
(iii) no real roots, no horizons. The two cases with one or two horizons are 
the interesting ones for the present analysis.
 
Let us now analyse other properties of the solution generated by the mass 
function \eqref{m2}. 

The Ricci scalar is obtained by replacing Eq.~\eqref{m2} into 
Eq.~(\ref{R1}), 
\begin{eqnarray}
R(r)=-\frac{q^4}{2mr^5}e^{-q^2/(2mr)}\label{R2}\,.
\end{eqnarray}
The Ricci scalar is negative and has a minimum value given by 
$R_{min}=-(50000\, m^4)/(e^5q^6)$ (with $e$ standing for the Euler number),
which happens at $r=q^2/(10 m)$.
We may invert expression \eqref{R2} to get  
\begin{eqnarray}
r(R)=-\frac{q^2}{10m}W^{-1}\left[-\frac{1}{5} 
\left(\frac{-q^6R}{(2m)^4}\right)^{1/5}\right]\label{rR1}.
\end{eqnarray}
where we have chosen the branch of the  Lambert function that produces real 
non-negative values for $r(R)$. The minus sign that appears in front of 
variable $R$ in the argument of the Lambert function guarantees we take the 
real and positive solutions for $r(R)$ because the Ricci scalar is negative for 
all $r$. It is worth mentioning that $R(r)$ given by relation~\eqref{R2} is not 
a one to one function of $r$, indeed it is a double valued function of $r$, and 
so its inverse, the Lambert function $W=r(R)$, must be dealt with care. 
Different branches of the Lambert function must be chosen to recover the whole 
range of the radial coordinate $r$. Namely, $r=W_0(R)$ to recover the interval 
$q^2/10m \leq r <\infty$ and $r=W_{-1}(R)$ to recover the interval $0\leq 
r<q^2/10m$. 
Further difficulties may appear in the points where the Ricci scalar vanishes, 
namely, at the center $r=0$ and at infinity $r\rightarrow\infty$. In these 
points the analysis has to be made separately (see Sec. \ref{secA1}
of Appendix \ref{secA}).

Then, substituting (\ref{rR1}) into (\ref{f1}) and integrating the resulting 
expression we get   
\begin{eqnarray}
&&\hspace{-.9cm} f(R)=c_0R-c_1\frac{q^2R}{6250m}W^{-5}\! 
\left[-\frac{1}{5}\!\left(\frac{-q^6R}{(2m)^4}\right)^{\!1/5}\right]
\Bigg\{ -6 +625W^4\left[-\frac{1}{5}\!\left(\frac{-q^6R}{(2m)^4}\right)^{\!1/5 
} \right ] \nonumber\\
&&
 +125W^3\left[-\frac{1}{5}\!\left(\frac{-q^6R}{(2m)^4}\right)^{\!1/5}\right]
-75W^2\left[-\frac{1}{5}\!\left(\frac{-q^6R}{(2m)^4}\right)^{\!1/5}\right]
+30W\!\left[-\frac{1}{5} 
\!\left(\frac{-q^6R}{(2m)^4}\right)^{\!1/5}\right]\!\Bigg\}
\label{f3} .
\end{eqnarray}
Taking the derivative of the last expression with respect to $R$ we 
find \begin{eqnarray}
f_R(R)=c_0-\frac{c_1q^2}{10m}W^{-1} 
\left[-\frac{1}{5}\left(\frac{-q^6R}{(2m)^4}\right)^{1/5}\right]\label{fR2}.
\end{eqnarray}
The behavior of functions $f(R)$ and $f_R(R)$ is shown in Fig.~\ref{ffR} but 
before analyzing it we present the analytical expressions of the other 
interesting functions.

Let us now turn attention to the electromagnetic source. The nonzero component 
of the Faraday-Maxwell tensor follows by replacing $M(r)$ from Eq. 
(\ref{m2}) into Eq. (\ref{F10-2}). The result is  
\begin{eqnarray}
F^{10}(r)& = &-\frac{e^{-q^2/(2mr)}}{4\kappa^2 mq
r^3}\Big\{c_1r\left[q^2(q^2-6mr)-4mr^2 
\left(3m-re^{q^2/(2mr)}\right)\right]\nonumber \\
&&+c_0q^2(q^2-8mr)\Big\}\label{F10-3} .
\end{eqnarray}
The final step is to build the Lagrangian density $\mathcal{L}_{NED}(F)$ in 
terms of the invariant $F$. Since there is not a closed 
(analytical) expression for that functional, we will show its behavior by 
means of a parametric plot (see Fig.~\ref{fig3}). 

\begin{figure}[h]
\centering
\begin{tabular}{rl}
\includegraphics[height=4.cm,width=6cm]{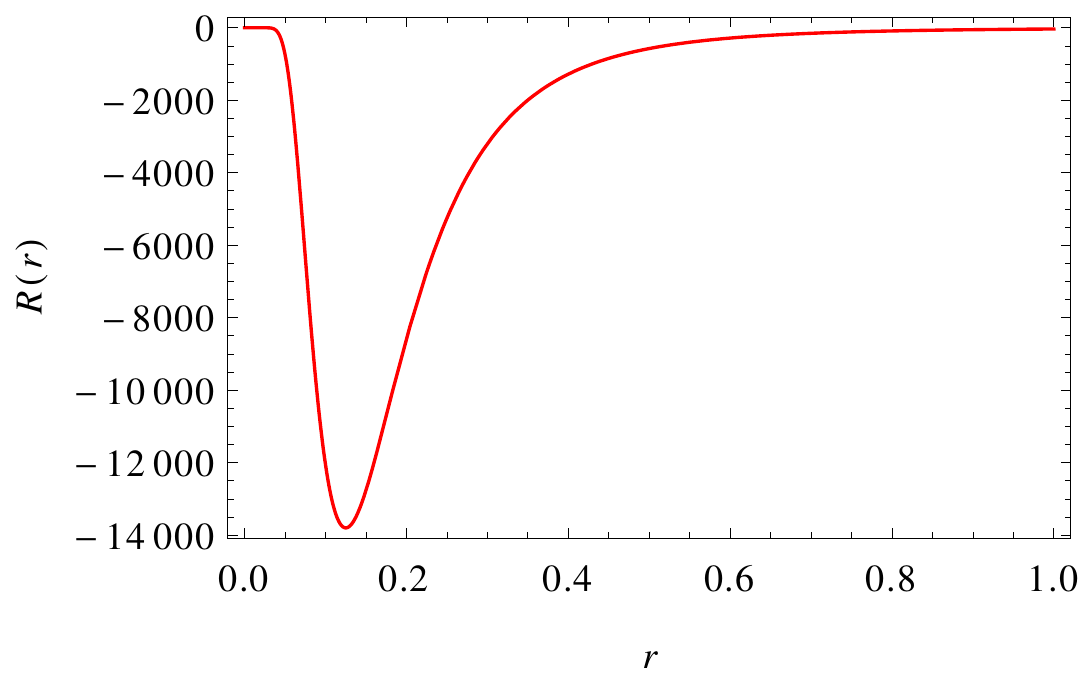} \hspace*{.5cm}
&\includegraphics[height=4.cm,width=6cm]{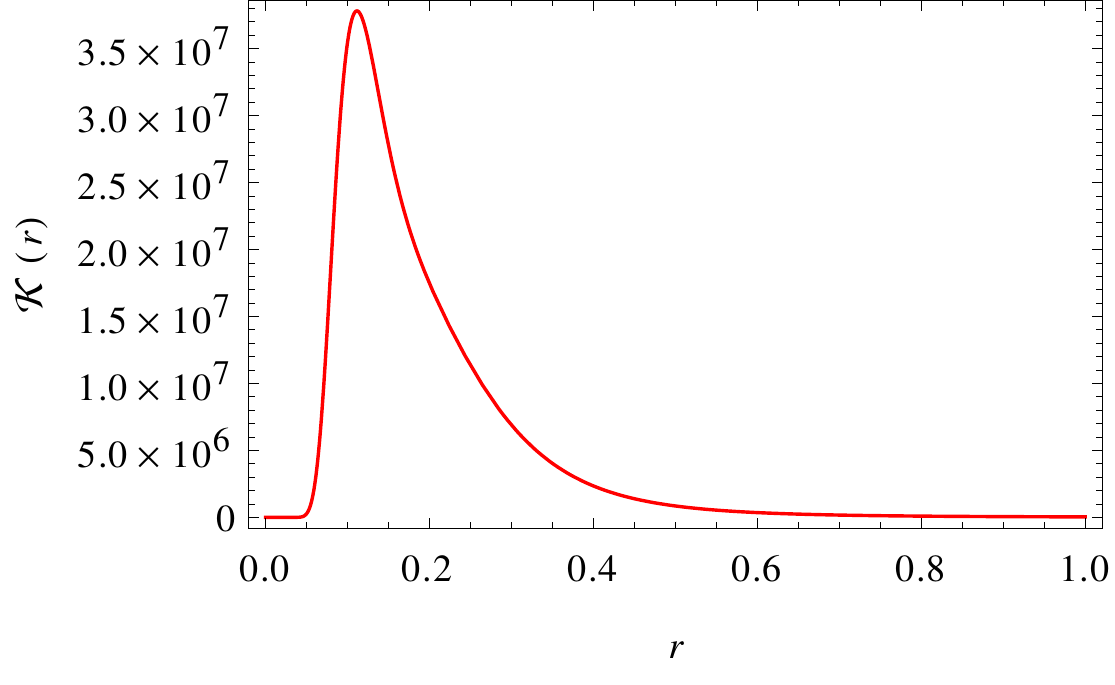}
\end{tabular}
\caption{{Graphical representation of functions $R(r)$ (left) 
and  ${\cal K}(r)$ (right), in the nondegenerated case, for 
the 
values $\{m=80,q=10,c_0=1,c_1=2,\kappa^2=8\pi\}$ of the solution 
(\ref{a1}). The minimum value of the Ricci scalar is $R_{min}=-13,799.3$, 
and is located at $r=1/8$. The maximum of the Kretschmann scalar is
${\cal K}_{max}=3.78138\times10^7$, and it happens at $r=0.11137$.}} 
\label{fig2}
\end{figure}

\begin{figure}[h]
\centering
\begin{tabular}{rl}
\includegraphics[height=4.0cm,width=6.0cm]{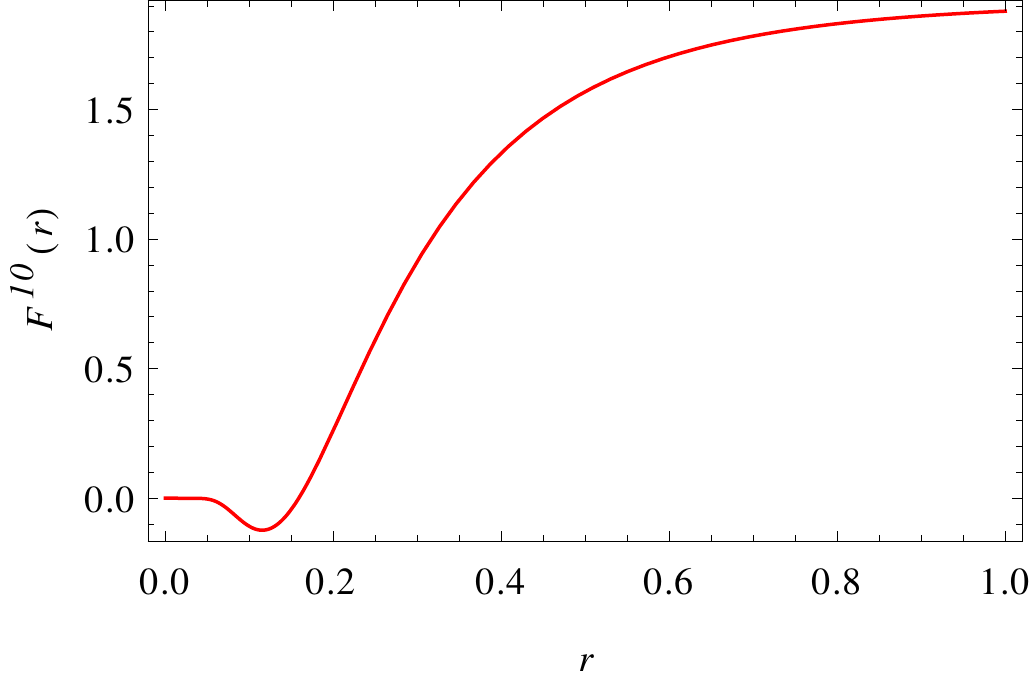} \hspace{.7cm}
&\includegraphics[height=4.0cm,width=6.0cm]{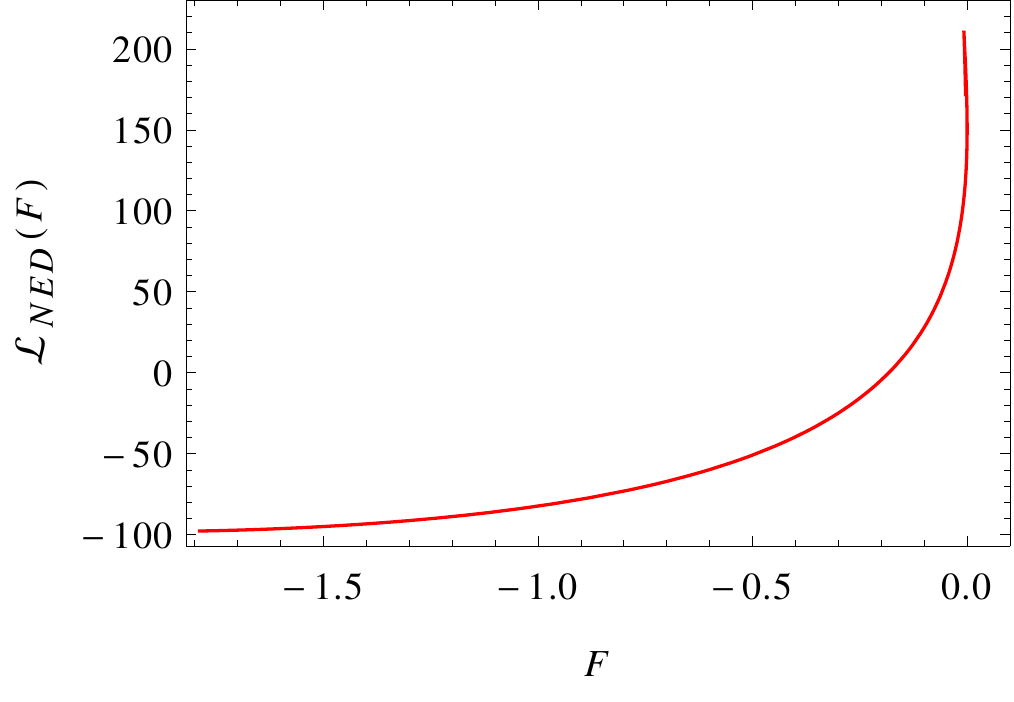}
\end{tabular}
\caption{{Graphical representation of functions $F^{10}(r)$ 
(left) and $\mathcal{L}_{NED}(F)$ (right), in the 
nondegenerated case, for the values 
$\{m=80,q=10,c_0=1,c_1=2,\kappa^2=8\pi\}$ of the solution (\ref{a1}).}} 
\label{fig3}
\end{figure}

\begin{figure}[h]  
\centering
\begin{tabular}{rl}
\includegraphics[height=4cm,width=6cm]{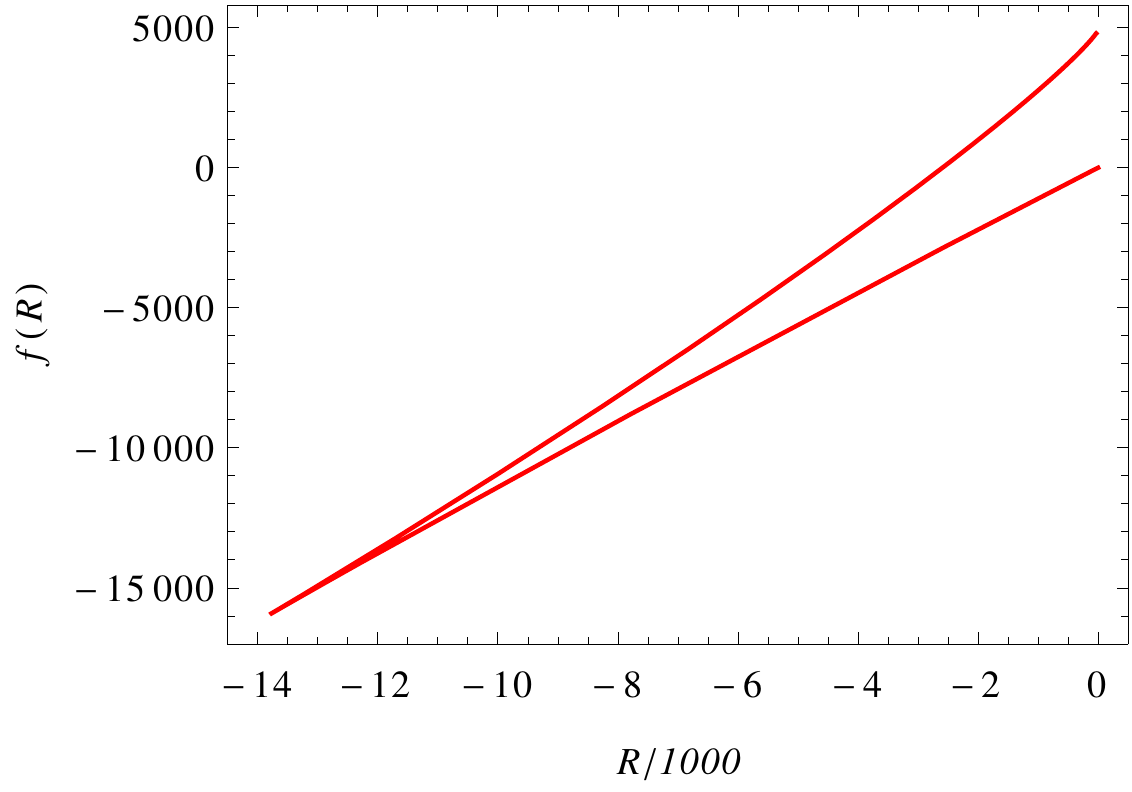} \hspace*{.5cm}
&\includegraphics[height=4cm,width=6cm]{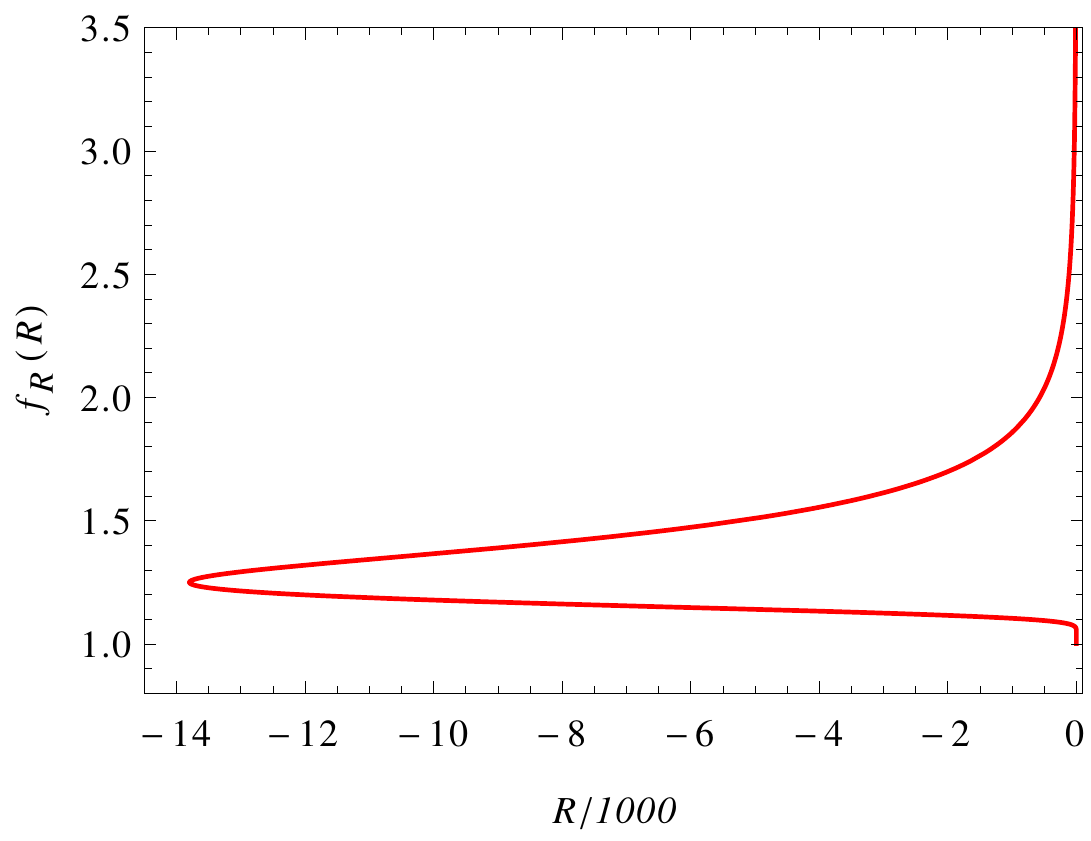}
\end{tabular}
\caption{{Graphical representation of functions $f(R)$ (left),  
and the function $f_R(R)$ (right), in the nondegenerated case, for the 
values $\{m=80,q=10,c_0=1,c_1=2,\kappa^2=8\pi\}$ of the solution 
(\ref{a1}). Both functions are defined, i.e., are real functions in the interval 
$R\in (0,-13,799.3] $.} \label{ffR}  } 
\label{fig4}
\end{figure}

Let us then analyse further the 
solution through numerical calculations.
First, it is seen that the solution is regular. In fact, in order 
to test for curvature singularities in a spacetime described by a metric of 
the form \eqref{ele}, it is sufficient to analyse the Ricci and Kretschmann 
scalars.
The Ricci scalar $R(r)$ given in  Eq.~\eqref{R2} is clearly regular 
throughout the spacetime. A graphic of $R(r)$ is shown in the left panel of 
Fig.~\ref{fig2}. 
The Kretschmann scalar for the solution (\ref{a1}) is given by
\begin{eqnarray}
&&\hspace{-.9cm} 
\mathcal{K}(r)=R^{\alpha\beta\mu\nu}R_{\alpha\beta\mu\nu}= 
\frac{e^{-q^2/(mr)}}{4m^2r^{10}}\!\big[
q^8-16mq^6r+96m^2q^4r^2-192m^3q^2r^3+192m^4r^4\big]\label{kre1}.
\end{eqnarray}
We represent graphically the Kretschmann scalar in the right panel of 
Fig.~\ref{fig2}.
As we can see, the present solution is asymptotically flat and regular at 
space infinity. One has $\lim_{r\rightarrow\infty} \{e^a(r),\, e^b(r)\}= 
\{1,\, 1\}$ and also $\lim_{r\rightarrow\infty} \{R(r),\,\mathcal{K}(r)\}= 
\{0,\, 0\}$. The solution is clearly regular for finite nonzero values of 
$r$. Furthermore, the solution is also regular at the origin of the radial 
coordinate. One has $\lim_{r\rightarrow 0} \{e^a(r),\, e^b(r)\}= 
\{1,\, 1\}$ and also $\lim_{r\rightarrow 0} \{R(r),\,\mathcal{K}(r)\}= 
\{0,\, 0\}$. This shows explicitly the regularity of the solution. 

Other interesting feature of the present solution is that it is a 
generalization of the solution presented in Eq. (17) of Ref.~\cite{balart}.
This is seen by taking the particular case $c_1 = 0$, in which case our 
Eqs.~(\ref{f3}) and (\ref{fR2}) turn into the GR corresponding functions. 
Moreover, for $c_1=0$ and $c_0=\kappa^2 /2$, the electric intensity
(\ref{F10-3}) reduces to Eq. (19) of 
Ref.~\cite{balart}. Function $F^{10}(r)$ is plot in Fig.~\ref{fig3} (left 
panel), where it is seen the regularity of such a function at the central 
core. The asymptotic behavior for large $r$ of this and related functions 
is investigated in detail in Sec.~\ref{secA1} of Appendix \ref{secA}.

A parametric plot of the Lagrangian density $\mathcal{L}_{NED}(F)$ in terms 
of the scalar $F$ is shown in the right panel of Fig.~\ref{fig3}. The 
nonlinear character of such a relation becomes evident.

Functions $f(R)$ and its derivative $f_R(R)$, given respectively by Eqs. 
(\ref{f3}) 
and (\ref{fR2}), are represented graphically in Fig. \ref{fig4} for a
particular choice of parameters. The nonlinear character of these curves 
reflects the fact that the gravity theory is not general relativity. It is 
clear they are well behaved functions at the central core.
More details on these functions, in particular, on the asymptotic 
behavior at the spatial infinity are given in Appendix \ref{secA}.  

In the regular black hole solution presented here, the metric functions  
$\{a(r),b(r)\}$, as well as the curvature scalars  
$\{R(r),\mathcal{K}(r)\}$, are identical to those presented in 
Ref.~\cite{balart} that considered GR coupled to NED. However, the 
new ingredient here, that guarantees we have a new solution, is exactly the 
function $f(R)$ given in Eq.~(\ref{f3}). It enters the action integral 
modifying the gravitational part of the Lagrangian density, and generalizes 
also the nonlinear Lagrangian density $\mathcal{L}_{NED}(F)$, which is 
modified because of the generalized Faraday-Maxwell tensor (\ref{F10-3}).

A further interesting analysis is to check if the present solution 
satisfies the energy conditions for the $f(R)$ gravity \cite{santos}.
Substituting the function $M(r)$ of the present model 
from Eq.~\eqref{m2} into the energy conditions (\ref{cond1})-(\ref{cond3}), 
and taking into account Eqs. (\ref{rhoeff})-(\ref{pteff}), we find
\begin{eqnarray}
NEC_1(r)&=&WEC_1(r)=0,\\ 
NEC_2(r)&=&WEC_2(r)=  \frac{q^2e^{-q^2/(2mr)}}{4m\kappa^2 
r^5}\left(8mr-q^2\right), \\
SEC(r)&=&\frac{q^2e^{-q^2/(2mr)}}{2m\kappa^2 r^5}\left(4mr-q^2\right), \\ 
DEC_1(r)&=& \frac12 DEC_2(r)=
4\times DEC_3(r)=   \frac{q^2}{\kappa^2 
r^4}e^{-q^2/(2mr)}\label{a1cond2}\;.
\end{eqnarray}
Therefore, the energy conditions are all satisfied in the region $r\geq 
(q^2/4m)$, a lower bound on $r$ imposed by the strong energy condition, 
$SEC(r)\geq 0$. The weak energy condition is satisfied if $r\geq (q^2/8m)$, a 
lower bound on $r$ imposed by $WEC_2(r)\geq 0$. For the choices 
of parameters as those chosen for drawing the graphs of 
Figs.~\ref{fig2}-\ref{fig4}, namely $\{q=10,m=80\}$, it follows that the 
$SEC$ is violated for  $r<0.3125$, while the $WEC$ is violated in the region 
$r<0.15625$. 
The region where these energy conditions are not satisfied is well inside 
the 
event horizon, since here we have $r_H=159.374$. It is well known that the 
$SEC$ is violated inside the horizon for the regular black hole solutions in 
GR theory, while the $WEC$ is in some cases violated throughout the 
spacetime.
Restricting the preset solution (\ref{a1}) to the GR theory, i.e., taking 
$\{c_0=1,c_1=0\}$, we see that both the $SEC$ and the $WEC$ are not 
satisfied in the same regions as for the $f(R)$ gravity theory. This is 
easily explained by noticing that the energy conditions 
(\ref{cond1})-(\ref{cond3}) do not depend on the parameters $c_0$ and $c_1$, 
and, moreover, the contributions from the functions $f(R)$ and $f_R$ to the 
effective energy density and pressures in
Eqs.~(\ref{rhoeff})-(\ref{pteff}) are mutually canceled.

In the next section we shall present a new solution which violates only the 
$SEC$ in a limited region of the spacetime.

\subsection{Second new regular black hole solution}\label{sec3.4}
Consider now the following ansatz for the mass function,
\begin{eqnarray}
M(r)=m\left(1+\frac{q^2}{4\beta mr}\right)^{-2\beta}\label{m3},
\end{eqnarray}
where $m$, $q$, and $\beta$ are constant parameters.
Such a model has been studied in Ref.~\cite{balart} in the context 
of GR coupled to NED theory. In GR it yields a black hole solution that is 
regular for $\beta\geq 3/2$ and that also satisfies the $WEC$ for $\beta\leq 
3/2$. 
Hence, in  the case where $\beta=3/2$ it results in a regular black 
hole which does satisfy the $WEC$. Since $\beta=3/2$ is a boundary value for 
regularity and to satisfy the $WEC$ within GR, we choose exactly this case to 
be analyzed within $f(R)$ gravity. Following the same steps as in the 
preceding section, we obtain 
\begin{eqnarray}
&&\hspace{-.5cm}e^{a(r)}=e^{-b(r)}=1-\frac{432m^4r^2}{(q^2+6mr)^3},\label{a2}\\
&&\hspace{-.5cm} \mathcal{K}
(r)=\frac{4478976m^8}{\left(q^2+6mr\right)^{10}}
\left(q^8+126m^2q^4r^2-216m^3q^2r^3+648m^4r^4\right) \label{kre2}\,,\\
&&\hspace{-.5cm}R(r)=-\frac{5184m^4q^4}{(q^2+6mr)^5}, \label{R(r)2}\\
&&\hspace{-.5cm} F^{10}(r)=\frac{(6mr)^3}{\kappa^2\left(q^2+6mr\right)^5} 
\Bigg\{72c_0m^2q +\frac{c_1}{q}\left[18m^2r(5q^2+6mr)-  
\frac{(q^2+6mr)^5}{216m^3r^2}\right]\Bigg\}\label{F10-4}.
\end{eqnarray}
The present solution is asymptotically flat and regular at 
spatial infinity. One has $\displaystyle{\lim_{r\rightarrow\infty} \\
\{e^a(r),\, e^b(r)\}= 
\{1,\, 1\}}$ and also $\displaystyle{\lim_{r\rightarrow\infty} 
\{R(r),\,\mathcal{K}(r)\}= 
\{0,\, 0\}}$. The solution is also regular at the origin of the radial
coordinate. One has $\displaystyle{\lim_{r\rightarrow 0} \{e^a(r),\, e^b(r)\}= 
\{1,\, 1\}}$
and also $\displaystyle{\lim_{r\rightarrow 0} \{R(r),\,\mathcal{K}(r)\}= 
\{-5184m^4/q^6,\,
4478976m^8/q^{12}\}}$. This shows explicitly the regularity of the solution.
\par 
We may use Eq.~\eqref{R(r)2} to get $r$ as a function of $R$ what, after 
being substituted into Eq.~(\ref{fR1}) and integrated, furnishes,  
\begin{eqnarray}
f(R)& = &c_0R+\frac{216m^3c_1}{q^4} 
-\frac{c_1}{12m}\left[2q^2R+5(162m^4q^4)^{1/5}R^{ 4/5}\right].\label{f4}
\end{eqnarray}
The present solution for $f(R)$ shows clearly the deviation from GR due to 
the presence of the last term in Eq. \eqref{f4}.
If we take $c_1=0$ and also $c_0=\kappa^2/2$ the GR theory is 
recovered. Notice also that in such a case the electric intensity 
(\ref{F10-4}) is 
identical to Eq. (49) of Ref.~\cite{balart}. The regularity of the  
functions at the central core is clear, while their behavior for large $r$ is 
investigated in detail in Sec. \ref{secA2} of Appendix \ref{secA}.

\begin{figure}[h]
\centering
\begin{tabular}{rl}
\includegraphics[height=4.5cm,width=6.7cm]{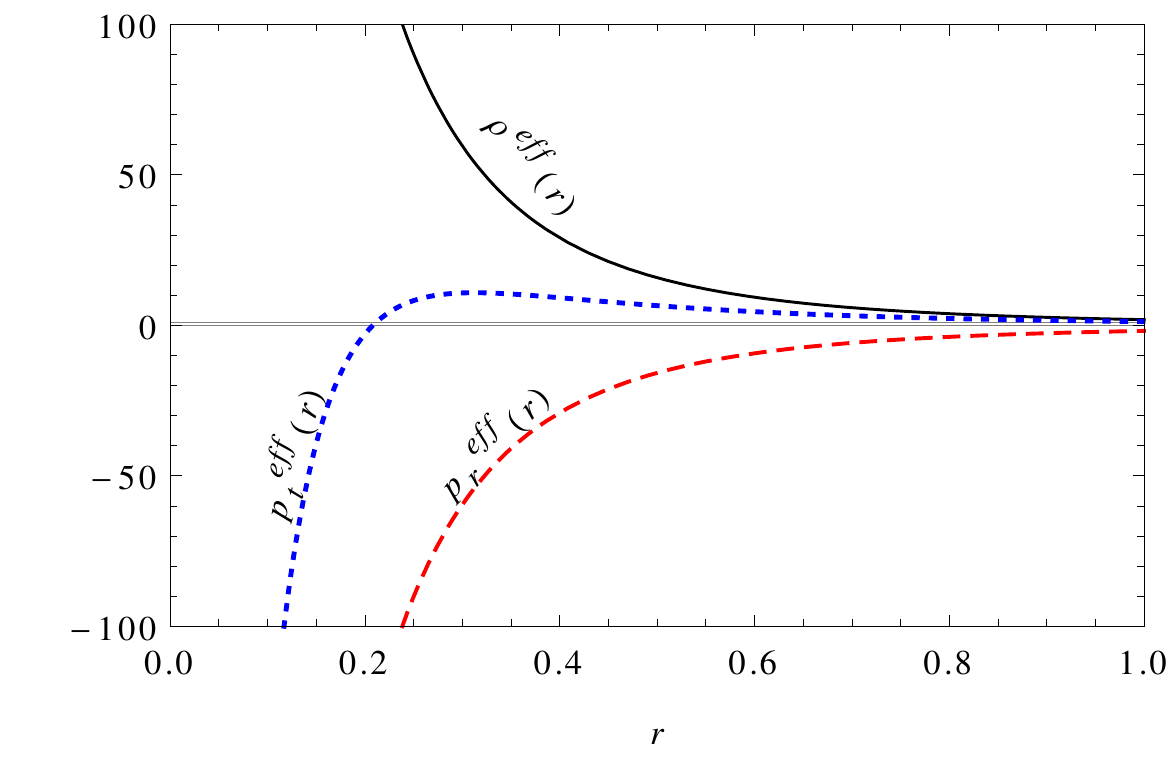} \hspace*{.5cm}
&\includegraphics[height=4.5cm,width=6.7cm]{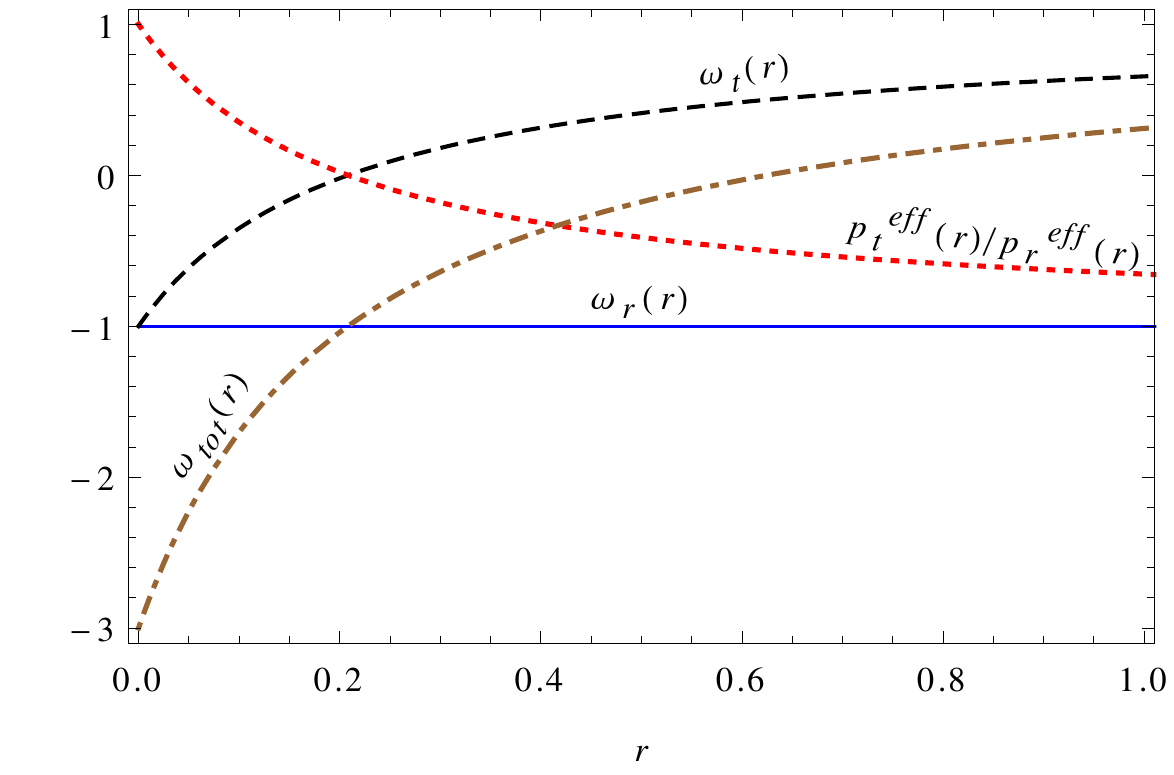}
\end{tabular}
\caption{{Graphical representation of the effective fluid 
quantities (left), and the functions  $\omega_r = p_r^{(eff)}/\rho^{(eff)}$, 
$\omega_t= p_t^{(eff)}/\rho^{(eff)}$, $p_t^{(eff)}/ p_r^{(eff)} $, and
$\omega_{tot} = (p_r^{(eff)} + 2 p_t^{(eff)})/\rho^{(eff)}$ (right), in the 
nondegenerated case, for the 
values $\{m=80,q=10,c_0=1,c_1=2,\kappa^2=8\pi\}$ of the solution  
    (\ref{m3}).}} 
\label{fig5}
\end{figure}

Here we analyse the energy conditions for the present solution. First we 
calculate the effective fluid quantities $\rho^{(eff)}$, $p_r^{(eff)}$ and 
$p_t^{(eff)}$ respectively from Eqs.~(\ref{rhoeff})-(\ref{pteff}). The 
result is
\begin{eqnarray} 
& & \rho^{(eff)} (r) = - p_r^{(eff)}(r) = 
\frac{1296\,q^2\,m^4}{\kappa^2\left(q^2 + 6m\,r\right)^{4}}
\label{rhoeff2},\\
& & p_t^{(eff)}(r)=   
\frac{-q^2 +6m\,r} {q^2 + 6 m\,r\,} \rho^{(eff)} (r).\label{pteff2} 
\end{eqnarray} 
The behavior of each one of these functions in terms of the radial 
coordinate $r$ is shown in the left panel of Fig. \ref{fig5}. The radial 
pressure satisfies the relation $p_r^{(eff)} =-\rho^{(eff)}$, while the 
tangential pressure is positive for large $r$ and tends to $-\rho^{(eff)}$ 
at the center, where the effective fluid behaves approximately as an  
isotropic fluid satisfying a de Sitter equation of state, $p^{(eff)} 
\simeq -\rho^{(eff)}$. 

The right panel of Fig. \ref{fig5} shows the ratios between the fluid 
quantities $\omega_r = p_r^{(eff)}/\rho^{(eff)}$ (solid line), 
$\omega_t= p_t^{(eff)}/\rho^{(eff)}$ (dashed line), and $p_t^{(eff)}/ 
p_r^{(eff)} $ (dotted line).
 We plot also the ratio $\omega_{tot}=  (p_r^{(eff)} + 2 
p_t^{(eff)})/\rho^{(eff)}$ (dot-dashed line). It is seen that the ratio 
$\omega_r$ is constant, $\omega_r=-1$. The ratio $\omega_t$ varies from 
$-1$ 
(for $r=0$) to $1$ (for $r\rightarrow\infty)$. Similarly we get $-1\leq 
p_t^{(eff)}/ p_r^{(eff)} \leq 1$, and $-3\leq \omega_{tot} \leq 1$.

Now, the energy conditions are found by substituting Eqs. 
(\ref{rhoeff2})-(\ref{pteff2}) into Eqs.~(\ref{cond1})-(\ref{cond3}),
\begin{eqnarray}
NEC_1(r)&=&WEC_1(r)=0,NEC_2(r)=WEC_2(r)=\frac{15552m^5q^2r}{
\kappa^2(q^2+6mr)^5}\label{cond4},\\
SEC(r)&=&\frac{2592m^4q^2(6mr-q^2)}{\kappa^2(q^2+6mr)^5},\label{cond5}\\
DEC_1(r)&=&\frac 12 DEC_2(r)=\frac{1296m^4q^2}{\kappa^2(q^2+6mr)^4},
 \nonumber\\
DEC_3(r)&=&\frac{2592m^4q^4}{\kappa^2(q^2+6mr)^5}\label{cond6}\;.
\end{eqnarray} 
The $NEC$, the $WEC$, and the $DEC$ are satisfied in the whole spacetime. 
On the other hand, the $SEC$ is not satisfied in the region $r< q^2/6m$. 
Once more, the $SEC$ is violated 
only in the central core, inside the event horizon. For the present case 
with $\{q=10,m=80\}$ it follows $q^2/6m = 5/24\simeq 0.208333$, while the 
horizon is at $r_H=159.374$. In conclusion, the only violated energy 
condition is the $SEC$, reproducing essentially the same result as the 
corresponding solutions in GR coupled to NED theory. Again, we can see that 
the final expressions for the energy 
conditions do not depend on the parameters of the $f(R)$ gravity, i.e., they 
do not depend on $c_0$ and $c_1$ and, because of that, the results are the 
same in both theories.

\section{Conclusion}\label{sec4}
We presented a class of solutions to the $f(R)$ gravity that generalizes 
 some regular black hole solutions found in GR theory. The energy-momentum 
tensor comes from a nonlinear electrodynamics theory minimally coupled to 
the $f(R)$ gravity. We have shown explicitly two solutions that are regular 
throughout the spacetime. The metric of the two solutions approach the
Reissner-Nordstr\"om 
metric far from the center, and possess also two free 
parameters corresponding to the ADM mass $m$ and to the electric charge 
$q$, respectively.

There is a third free parameter, $c_1$, related to the $f(R)$ function, such 
that the $f(R)$ theory reduces to general relativity when such a parameter 
vanishes.
Moreover, by writing the resulting field equations for the $f(R)$ gravity in 
terms of the Einstein tensor, an effective energy-momentum tensor can be 
defined.
The result is an anisotropic perfect fluid. The effective energy density, 
radial and tangent pressures for the two new solutions analyzed do not depend 
on the 
parameter $c_1$, suggesting that the properties of the matter-energy content 
of these solutions are similar to the corresponding geometries in general 
relativity theory.

In fact, the first solution violates the null and the weak 
energy conditions at the central region, well inside the event horizon, 
whose size is given by the radial coordinate characterized by the ratio 
$r_0=
q^2/8m$.  These conditions are violated in the region inside $r_0$, 
$r<q^2/8m$.
It also violates the strong energy condition in a spherical region $r< 
q^2/4m$.
The second of those solutions satisfies all but the strong energy condition 
that is violated in a central core well inside the event horizon, with $r< 
q^2/6m$. These results are in agreement with the singularity theorems of 
general relativity (see also \cite{zaslavskii}).

The regularity of the solutions is verified through the Ricci and Kretschmann 
scalars, assuring that there are no spacetime (curvature) singularities. The 
asymptotic 
behavior of the physical quantities at the spatial infinity is analyzed in 
view of the fact that two particular quantities, namely the derivative of the 
gravity Lagrangian density $df(R)/dR$ and the electric intensity $F^{10}$,
diverge linearly at the limit $r\rightarrow\infty$. We show 
explicitly that all the relevant physical quantities for both of the 
solutions are bounded and well behaved in the asymptotic limit.

As a further development in this line of investigation, we expect that the 
method employed here may be used to obtain new 
interesting solutions, not only for regular black holes, generalizing those 
obtained in the context of general relativity. A more careful and detailed 
analysis related to the energy conditions in $f(R)$ seems necessary and, in 
fact, is one of the subjects under investigation by ourselves.

The present line of work also opens the possibility of new applications 
to local astrophysical phenomena such as the dark matter problem. For 
instance, a phenomenologically motivated mass function $M(r)$ may be 
proposed to fit the rotation curves of galaxies within some specific $f(R)$ 
model. Other application of the strategy followed in the present work 
is to simulate the mass distribution at the galactic bulge for a specific 
galaxy or some galaxy sample.

\vspace{1cm}

{\bf Acknowledgements}: M. E. R.  
thanks Conselho  
Nacional de Desenvolvimento Cient\'ifico e Tecnol\'ogico - CNPq, Brazil, 
Edital MCTI/CNPQ/Universal 14/2014  for partial financial support. 
V. T. Z.  thanks Funda\c c\~ao de
Amparo \`a Pesquisa do Estado de S\~ao Paulo (FAPESP), Grant No. 
2011/18729-1, 
Conselho Nacional de Desenvolvimento Cient\'\i fico e Tecnol\'ogico of 
Brazil (CNPq), Grant No. 308346/2015-7, and Coordena\c{c}\~ao de
Aperfei\c{c}oamento do Pessoal de N\'\i vel Superior (CAPES), Brazil, Grant
No.~88881.064999/2014-01.

\appendix
\section{The asymptotic limit of the two new solutions}
\label{secA}
Here we explore the asymptotic limit $r\rightarrow +\infty$ of the two
solutions presented above. The aim is to verify that the solutions are
regular in such a limit. We have already commented in the text that all the
physical quantities related to those solutions are regular everywhere in
the spacetime. However, since some of the functions assume arbitrarily
large values at $r\rightarrow\infty$ it is worth dwelling longer on this
subject. To simplify analysis we define $r=1/x$ and then write all the
functions in terms of the new variable $x$. The asymptotic limit of interest
is now $x\rightarrow 0$.

\subsection{Analysis of the first solution}
\label{secA1}
Let us start with the metric function $e^{a(r)}$ of the first solution
discussed in Sec.~\ref{sec3.3}. From Eq.~\eqref{a1},we get
\begin{eqnarray} e^{a(r)}= e^{a(1/x)}= 1-2mx+q^2x^2+O\left(x^3\right).
\label{A1}
\end{eqnarray}  
After the hypothesis \eqref{bconstraint}, $e^{b(r)}=e^{-a(r)}$, the last
equation implies the metric functions of the present solution approach the
Reissner-Nordstr\"om metric up to the second order in $1/r$. 

On the other hand, function $df/dR$ given by Eq.~\eqref{fR1} is clearly
divergent at the radial infinity. However, what matters here is to verify
whether such a divergence implies any kind of inconsistency in the physical
and geometrical quantities regarding the solutions obtained in the present
work. In fact, the Lagrangian density of the $f(R)$ gravity theory is the 
function $f(R)$ itself and, for the present case, it is given by 
Eq.~\eqref{f3}. Expanding such a function
in powers of $x=1/r$ one gets
\begin{eqnarray}
f(R)=\frac{48m^3}{q^4}c_1-\frac{5q^4}{8m}c_1x^4+O\left(x^5\right)
\label{A2}.
\end{eqnarray}  
Hence, the gravity Lagrangian density tends to a constant which means no 
diverge problems
at the radial infinity. 

Let us now analyse the functions related to the nonlinear electrodynamics 
theory employed here.
The starting point is the Lagrangian $\mathcal{L}_{NED}$, whose
asymptotic form, obtained from Eq. \eqref{L}, for the model defined by
Eq.~\eqref{m2}, is
\begin{eqnarray}
\mathcal{L}_{NED}=
-\frac{24m^3}{q^4\kappa^2}c_1-\frac{c_1}{\kappa^2}x+O\left(x^3\right)
\label{A3}, 
\end{eqnarray}
indicating that the NED Lagrangian density is also well behaved at the
asymptotic limit. We see that the Lagrangian
$\mathcal{L}_{NED}$ tends to a constant which exactly cancels the 
contribution from the gravity $f(R)$ Lagrangian density at first-order 
approximation, leading to a vanishing total Lagrangian density at the 
asymptotic limit, as usual for asymptotically flat spacetimes.

Now, taking the approximate form  for the derivative of the NED Lagrangian
density with respect to the field $F$, from \eqref{LF}, we get 
\begin{eqnarray}
\mathcal{L}_F= -\frac{q^2\kappa^2}{c_1}x^3+O\left(x^4\right)\label{A4},
\end{eqnarray}
which is also well defined at the asymptotic limit.
The asymptotic form of the field $F^{10}(r)$ for the present
solution is obtained from Eq.~\eqref{F10-3}, 
\begin{equation}
 q\kappa^2 F^{10} = 3\, c_1m-\frac{c_1}{ x} + \frac{q^2\left(16m\,c_0 - 
5q^2c_1 \right) x^2 } {8m} + O(x^3). \label{A4b}
\end{equation}
Even though this function diverges with $r$, the electromagnetic energy
density, defined by the right-hand side of Eq.~\eqref{eq1} (divided by
$\kappa^2$), $\rho_{EM} =\mathcal{L}_{NED}+ 
\frac{q^2}{r^4}\mathcal{L}_F^{-1}$, is well behaved at $r\rightarrow\infty$.
In fact, we have
\begin{eqnarray}
\rho_{EM} = \frac{24m^3}{q^4}c_1+2c_1x-3mc_1x^2+O\left(x^3\right)\label{A5},
\end{eqnarray}
which approaches a constant at the asymptotic limit. Furthermore,
the electric induction field $D^{\mu\nu} = \partial\mathcal{L}_{NED}/\partial 
F^{\mu\nu} =\mathcal{L}_F F^{\mu\nu} $ (see, e.g., \cite{burinskii}) is also 
well behaved at the spatial infinity. In fact, we obtain
\begin{equation}
 D^{10} =  \mathcal{L}_F\, F^{10} =  qx^2 - 3\,m\,q x^3 + 
O(x^4), \label{A4c}
\end{equation}
which vanishes according to $q/r^2$, as expected for a pointlike electric 
source in an asymptotically flat spacetime. As a matter of fact, this is a 
situation where the electric component of the Faraday-Maxwell tensor  
$F^{10}(r)$ does not fall-off as $1/r^2$ in an asymptotically Minkowskian 
spacetime (see, e.g., Ref.~\cite{gonzalezetal} for comparison with cases 
where the nonvanishing asymptotic behavior of $F^{10}(r)$ is connected to 
nonasymptotically flat spacetimes).

Moreover, we can verify that all the other physical quantities are also well 
behaved and regular
everywhere in the spacetime. An important quantity is the effective
energy density $\rho_{eff}$, given by Eq.~\eqref{rhoeff}. Its asymptotic
limit is
\begin{eqnarray}
\rho^{(eff)} = \frac{q^2}{\kappa^2}x^4+O\left(x^5\right)\label{A6},
\end{eqnarray}   
what is well behaved at infinity and guarantees that the total energy of
the solution is bounded. We have also performed the 
analysis of the effective pressures, $p_r^{(eff)}$ and $p_t^{(eff)}$, and 
verified that all the components of the effective energy-momentum tensor are 
well behaved functions at the asymptotic limit.

As a further consistency check, we investigate the behavior of the
equations of motion at $r\rightarrow +\infty$ ($x\rightarrow 0$).
After the choice \eqref{bconstraint} and, as a consequence, using the fact
that $d^2f_R/dr^2 =0$ (see Eq.~\eqref{fR2}), it follows that Eq.~\eqref{eq2}
is identical to Eq.~\eqref{eq1}, so that only Eqs.~\eqref{eq1} and
\eqref{eq3} are independent relations. 
In such a limit, the left-hand-side of Eq.~\eqref{eq1} is
\begin{eqnarray}
\frac{24m^3}{q^4}c_1+2c_1x-3mc_1x^2+O\left(x^3\right)\label{A7}, 
\end{eqnarray} 
showing no divergent term. The same result is found from the
right-hand side of that equation. 
In turn, the asymptotic limit of the left-hand side of \eqref{eq3} is
\begin{eqnarray}
\frac{24m^3}{q^4}c_1+c_1x+O\left(x^3\right)\label{A8},
\end{eqnarray}
which is the same as the right-hand side of that equation at the asymptotic
limit. 
These results show that there are no divergent term in the equations of
motion. Hence, even though the
function $f_R(r)$ is unbounded at $r\rightarrow\infty$ all the equations of
motion are well behaved everywhere.

At last we check the asymptotic form of the Ricci and Kretschmann scalars.
From Eqs.~\eqref{R2}  and \eqref{kre1} we get, respectively,
\begin{eqnarray}
&&R= -\frac{q^4}{2m}x^5+O\left(x^6\right)\label{A9},\\
&&\mathcal{K}= 48m^2x^6+O\left(x^7\right)\label{A10}.
\end{eqnarray}

We finish this analysis emphasizing that all the geometric invariants and 
physical quantities of the solution presented in Sec.~\ref{sec3.3} are 
regular, the equations of motions show no divergence and are well behaved 
everywhere in the spacetime. In this sense, we have found a fully consistent 
solution. The only unbounded functions are $f_R$ (see Eq.~\eqref{fR2}) and the 
nonlinear electric intensity $F^{10}$, given by Eq.~\eqref{F10-3}. The
divergence of $f_R$ does not imply any kind of divergence on the physical
and geometric 
properties of the spacetime. Moreover, the relevant physical 
quantities derived from the nonlinear electrodynamics are well behaved 
functions: the electromagnetic energy density is bounded, the energy-momentum 
tensor has no divergent terms, the total charge is finite (and constant), and 
the physical electric induction $D^{10} =
\partial\mathcal{L}_{NED}/\partial 
F^{10} =\mathcal{L}_F\,F^{10}  $ also behaves as expected for a pointlike 
electrostatic source.

\subsection{Analysis of the second solution}
\label{secA2}
Now we consider the asymptotic behavior of the solution derived from the 
ansatz of Eq.~\eqref{m3} [see also Eq.~\eqref{a2}] and presented in 
Sec.~\ref{sec3.4}. As above, we 
express the functions in terms of the new variable $x=1/r$.

As in the case of solution \eqref{a1} the asymptotic form of the resulting  
metric at $r\rightarrow\infty$ is the Reissner-Nordstr\"om metric, 
with 
$g_{tt}=1/g_{rr}\simeq 1-2mx+q^2x^2 - q^4x^3/3m$, where we fixed $\beta = 
3/2$.

The derivative of the gravitational Lagrangian density $f_R$ has the same 
form as in the case discussed in the last section, and so it diverges 
linearly with $r$. What we shall analyse once again is whether such a 
divergence introduces some inconsistency into the solution or not. First, 
using relation \eqref{f1} and \eqref{m3} we see that the Lagrangian density 
$f(R)$ is of the form 
\begin{eqnarray}
f(R) = 
\frac{216m^3}{q^4}c_1-\frac{5q^4}{6m}c_1x^4+O\left(x^5\right)\label{A11}, 
\end{eqnarray}  
which tends to a constant. 

The electromagnetic Lagrangian density $\mathcal{L}_{NED}$ also tends to a 
constant. In fact, using relations \eqref{L} and \eqref{m3} we find
\begin{eqnarray}
\mathcal{L}_{NED} = -\frac{108m^3}{q^4\kappa^2}c_1 
-\frac{c_1}{\kappa^2}x+O\left(x^3\right)\label{A12}.
\end{eqnarray}
Once again, we see that, as $f(R)$, the Lagrangian
$\mathcal{L}_{NED}$ is  well behaved at the spatial infinity. As in the 
case of Sex.~\ref{sec3.3}, the total Lagrangian density 
$2\kappa^2\mathcal{L}_{NED} + f(R)$ vanishes at the asymptotic limit, a 
common behavior in asymptotically flat spaces.

Taking the same limit of the derivative of the electromagnetic Lagrangian 
density $\mathcal{L}_F$, from Eqs.~\eqref{LF} and \eqref{a2} it results the 
same form as in Eq.~\eqref{A4}, showing that it is a bounded function in the 
asymptotic limit.

The asymptotic form of the nonzero components of the Faraday-Maxwell tensor
field $F^{10}(r)=-F^{01}(r)$ (the electric intensity) for the present 
solution
is obtained from 
E.~\eqref{F10-4}, \begin{equation}
 q\kappa^2 F^{10} (r) = 3\, c_1m-\frac{c_1}{ x} + \frac{q^2\left(12m\,c_0 - 
5q^2c_1
\right) x^2 } {6m} + O(x^3), \label{A12b}
\end{equation}
which has approximately the same form as in the first solution, and diverges 
linearly with $r$. However, the other electromagnetic quantities, the ones 
that have direct physical meaning, are well behaved function. 
For instance, the electromagnetic energy density  
\eqref{eq1} for the solution of section \ref{sec3.4}, at the asymptotic 
region, assumes the form
\begin{eqnarray}
\rho_{EM} = \frac{108m^3}{q^4}c_1+2c_1x-3mc_1x^2+O\left(x^3\right)\label{A13},
\end{eqnarray}
which tends to a constant. Moreover, the electric induction $D^{10}(r)$
 is also well behaved at the spatial infinity. Indeed, we find 
the same leading terms as in Eq.~\eqref{A4c}, and the electric induction
once again tends to $q/r^2$, as expected for a pointlike electric 
source in asymptotically flat spacetimes (see also the discussion following 
Eq.~\eqref{A4b} above).

Other interesting physical quantity, the effective energy density, is well 
behaved too. Its asymptotic form is obtained from Eq.~\eqref{rhoeff} [or 
from Eq.~\eqref{rhoeff2}] and the result is identical to the previous case, 
given by Eq.~\eqref{A6}. A similar analysis of the effective pressures shows 
that all the components of the effective energy-momentum tensor are bounded 
function at the asymptotic limit. 
 Another important consistency check is related to the equations of motion, 
since the divergence of functions  $f_R$ and $F^{10}$ may induce singularities 
into the equations. Considering the solution \eqref{a2} into Eq.~\eqref{eq1}, 
it follows that both sides of such an equation tend to 
\begin{eqnarray}
\frac{108m^3}{q^4}c_1+2c_1x-3mc_1x^2+O\left(x^3\right)\label{A14},
\end{eqnarray}
with no divergent terms and no other inconsistency is observed. As in the
case of the first solution, Eq.~\eqref{eq2} is identical to Eq.~\eqref{eq1}.
In turn, the left-hand side of Eq.~\eqref{eq3} gives
 \begin{eqnarray}
\frac{108m^3}{q^4}c_1+c_1x+O\left(x^3\right)\label{A15}, 
\end{eqnarray}
which is identical to the result obtained for the expression on the 
right-hand side of that equation. This shows there are no divergence nor 
inconsistencies in the equations of motions.

 At last, the asymptotic form of the Kretschmann and Ricci scalars, 
from Eqs.~\eqref{kre2} and \eqref{R(r)2}, respectively, are
\begin{eqnarray}
&&\mathcal{K}= 48m^2x^6+O\left(x^7\right)\label{A17}, \\
&&R= -\frac{2q^4}{3m}x^5+O\left(x^6\right)\label{A16}.
\end{eqnarray}
With this we complete the analysis and verify that there no inconsistencies 
neither divergences of the physical and geometric properties of the spacetime.


\end{document}